\documentclass[aps,prl,preprint,superscriptaddress,amsmath]{revtex4}

\usepackage{graphicx}
\usepackage{dcolumn}

\begin{document}

\title{Quantum Transport Length Scales in Silicon-based Semiconducting Nanowires:\\Surface Roughness Effects}
\author{Aur\'elien Lherbier}
%\email{aurelien.lherbier@cea.fr}
\affiliation{Laboratoire des Technologies de la Micro\'electronique (LTM), UMR 5129 CNRS, CEA 17 Rue des Martyrs 38054 Grenoble France}
\affiliation{CEA/DSM/DRFMC/SP2M/L\_Sim, 17 rue des Martyrs, 38054 Grenoble Cedex 9, France}
\author{Martin P. Persson}
\affiliation{CEA/DSM/DRFMC/SP2M/L\_Sim, 17 rue des Martyrs, 38054 Grenoble Cedex 9, France}
\author{Yann-Michel Niquet}
\affiliation{CEA/DSM/DRFMC/SP2M/L\_Sim, 17 rue des Martyrs, 38054 Grenoble Cedex 9, France}
\author{Fran\c cois Triozon}
\affiliation{CEA LETI-MINATEC, 17 rue des Martyrs, 38054 Grenoble, Cedex 9 France}
\author{Stephan Roche}
\affiliation{CEA/DSM/DRFMC/SPSMS/GT, 17 rue des Martyrs, 38054 Grenoble Cedex 9, France}

\date{\today}

\begin{abstract}
We report on a theoretical study of quantum charge transport in atomistic models of silicon nanowires with surface roughness-based disorder. Depending on the nanowires features (length, roughness profile) various conduction regimes are explored numerically by using efficient real space order $N$ computational approaches of both Kubo-Greenwood and Landauer-B\"uttiker transport frameworks. Quantitative estimations of the elastic mean free paths, charge mobilities and localization lengths are performed as a function of the correlation length of the surface roughness disorder. The obtained values for charge mobilities well compare with the experimental estimates of the most performant undoped nanowires. Further the limitations of the Thouless relationship between the mean free path and the localization length are outlined.
\end{abstract}

\maketitle

\section{INTRODUCTION}

Semiconducting silicon nanowires (SiNWs) are currently the subject of intense studies due to their prominent role in the downscaling limits of MOSFET devices, and also because they provide alternative materials to challenge quantum effects in low dimensionality \cite{Larsmagnetoresistance}. Compared to classical planar technology, nanowires can better accommodate "all-around" gates improving field effect efficiency and device performances \cite{GAR}. Vapor-liquid-solid (VLS) growth techniques have recently produced SiNWs with well controlled structural features \cite{Dick}, which opens up innovative approaches to the design of silicon-based nanodevices \cite{LARS}. Demonstrations of $p-n$ junction diodes \cite{YiCui02022001}, logic gates \cite{YuHuang11092001}, field effect transistors \cite{CuiY._nl025875l}, and nanosensors \cite{YiCui08172001} have been reported.

However, one key issue in the engineering of performant SiNW-based field effect transistors (SiNW-FETs) is to ascertain how sensitive the charge mobilities are to structural features such as diameter, growth direction, and disorder. Surface roughness disorder (SRD) is a well known limiting factor in lithographic SiNW-FETs \cite{Simob,Lundstrom}, and its impact on ballistic transport in VLS-grown nanowires is a challenging and important question \cite{CuiY._nl025875l}. Besides, SRD effects also raise fundamental questions in the framework of localization theory \cite{Juanjo,Feist}.

Recent {\it ab initio} studies \cite{PRL_Lorente,fernandez-serra:166805} have reported on specific surface effects, such as dopant segregation in small diameter SiNWs. However, these studies generally hardly cope with the analysis of the fundamental transport length scales in long disordered nanowires. Several theoretical works have also investigated the role of effective and simplified surface disorder models on the transport properties of nanowires-based materials or devices \cite{Troels,Zhong}.

In this work, we report on a quantitative analysis of the transport length scales in atomistic models of rough SiNWs. 
The description of the SiNWs is based on an accurate tight-binding Hamiltonian, previously validated by {\it ab initio} calculations \cite{niquet:165319}. 
The quantum transport properties are calculated with two different approaches. First, the elastic mean free path and the charge mobility are computed with an optimized, real-space order $N$ Kubo-Greenwood approach \cite{Kubo,triozon:121410}. Additionally, the scaling properties of the Landauer-B\"uttiker conductance are investigated with a standard recursive Green's function method, to assess the effects of quantum interferences driving to the localization regime. Both approaches give complementary results and allow to explore a broad range of conduction mechanisms, from the ballistic to the diffusive and strongly localized regimes.

\section{METHOD}

\subsection{Description of the Surface Roughness Profile}

The SiNW Hamiltonian is a third nearest neighbor three center orthogonal $sp^3$ tight-binding model that well describes the electronic structure of ideal (disorder-free) nanowires \cite{niquet:165319}. The SRD profile is defined as a random fluctuation of the radius of the nanowire around its average value $R_0$, characterized by a Lorentzian auto-correlation function, with a single intrinsic length scale \cite{Lundstrom,Ferry}. In cylindrical coordinates,

\begin{subequations}
\begin{equation}
\delta R(z,\theta) = \sum_{(n,k)\neq(0,0)} a_{nk} \: e^{in\theta} \: e^{i\frac{2\pi}{L}kz }  \label{EQ2}
\end{equation}
where:
\begin{equation}
a_{nk} = \frac{e^{i \varphi_{nk}}}{\left\lbrace 1+\left[\left( \frac{2\pi k}{L}  \right)^2 + \left( \frac{n}{R_0} \right)^2\right] L_r^2 \right\rbrace ^{3/4}}. \label{EQank}
\end{equation}
\end{subequations}

\begin{figure}[!t]
	\centering
	\includegraphics[width=8.5cm]{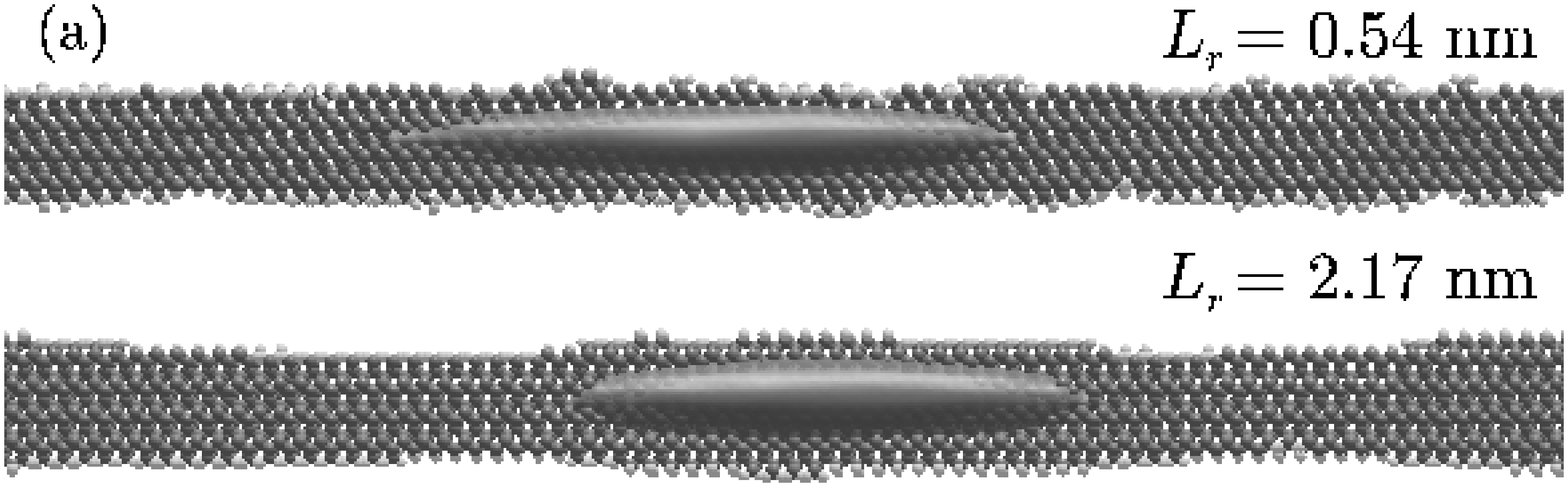}
	\includegraphics[width=8.5cm]{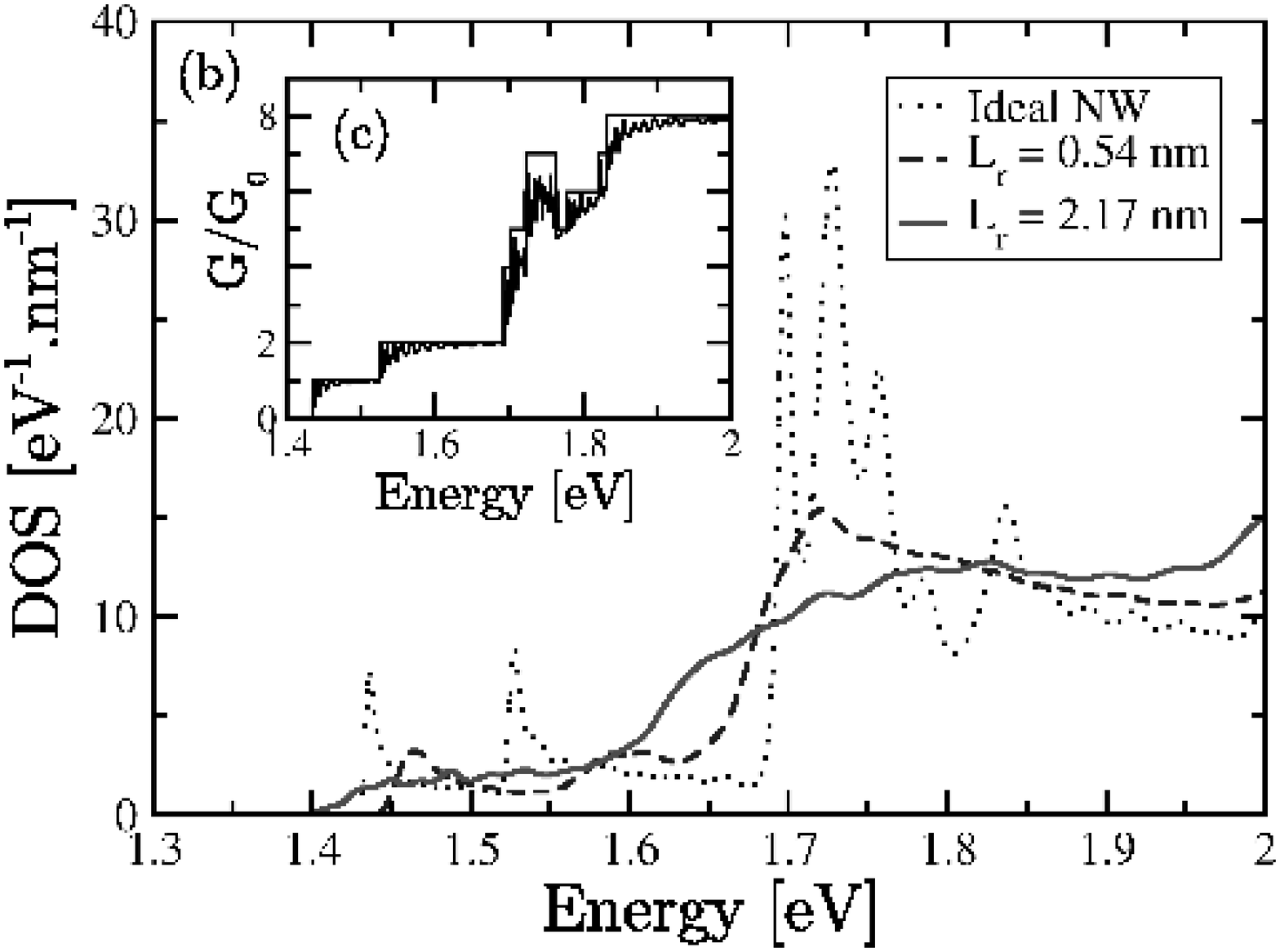}
	\caption{(color online). (a) SRD profile and isodensity surface of the ground-state conduction band wave function of small sections of $[110]$ SiNWs, for two values of $L_r$. (b) Corresponding conduction band DoS (dashed and solid lines). The DoS of the ideal nanowire with radius $R_0=1$ nm is also shown (dotted line). (c) Landauer-B\"uttiker conductance of the same ideal nanowire \cite{noteLB}, along with the number of conducting channels (staircase line).}
\label{figure1}
\end{figure}

$\varphi_{nk}\in[0,2\pi[$ is a random number, $L$ is the length of the nanowire and $L_r$ is the correlation length of the SRD. The silicon atoms outside the envelope defined by Eq. (\ref{EQ2}) are excluded from the nanowire and the dangling bonds are saturated with hydrogen atoms \cite{noteH}. In the following, we set $R_0=1$ nm and renormalize the $a_{nk}$'s so that $\left\langle \delta R^2(z,\theta) \right\rangle^{1/2} = 1$ \AA{}, leaving $L_r$ as the only free parameter. The effects of the SRD on the transport properties have been analyzed for [110] oriented SiNWs with $L_r$'s ranging from  0.54 to 4.34 nm [Fig.~\ref{figure1}(a)].

\subsection{Transport methodologies}

In the semiclassical transport theory, disorder effects can be characterized by the scattering rate between the eigenstates of the ideal system \cite{Mott}. The scattering time $\tau$ is usually computed with the Fermi's golden rule, and by virtue of Matthiessen's law, can be split into an elastic ($\tau_e$) plus an inelastic ($\tau_i$) contribution ($1/\tau = 1/\tau_e + 1/\tau_i$). The SRD is expected to dominate backscattering at low temperatures \cite{Simob}, while the inelastic electron-phonon coupling plays a major role at room temperature. In the Kubo-Greenwood approach, the scattering time $\tau_e(E)$ and the mean free path $\ell_{e}(E)=v(E)\tau_e(E)$ are extracted from the saturation of the quantum diffusivity $D(E,t)=\Delta Z^2 (E,t)/t\to 2v^2(E)\tau_e(E)$, where $v(E)$ is the average velocity and $\Delta Z^2 (E,t)$ the quadratic spreading of wave packets with energy $E$ \cite{triozon:121410}:
\begin{equation} 
\Delta Z^2 (E,t) = \frac{{\rm Tr}\left[\left[\hat{Z}(t)-\hat{Z}(0)\right]^2 \delta(E-\hat{H}) \right]}{{\rm Tr}\left[ \delta(E-\hat{H}) \right] }\,. \label{EQ1}
\end{equation}
$\hat{Z}(t)$ is the position operator in the Heisenberg representation, while $\delta(E-\hat{H})$ is the spectral measure of the SiNW Hamiltonian. Periodic boundary conditions are applied along the nanowire, the convergence being achieved for supercell lengths $L\simeq 500$ nm. The real-space methodology of Ref.~\cite{triozon:121410} has been adapted to this multi-orbital per site problem. In particular, we have used the kernel polynomial method \cite{KPM} to compute the spectral quantities from the Lanczos recursion coefficients \cite{0022-3719-17-22-013}. This provides a more accurate description of the band edges than the usual continued fraction expansion \cite{Troels,triozon:121410,noteavg}.

\section{RESULTS}

\subsection{Electronic properties}

Let us first discuss the electronic structure of the SiNWs. The conduction band density of states (DoS) for the ideal and for two disordered SiNWs are shown in Fig.~\ref{figure1}(b). In the ideal SiNW, the first two Van Hove singularities (VHs) arise from the $[001]$ bulk conduction band minima \cite{niquet:165319} and are split by the inter-valley couplings, while the others (above 1.7 eV) arise from both $[001]$ and $\{[100],[010]\}$ minima. The DoS is markedly affected by the SRD. At $L_r=0.54$ nm, the lowest-lying VHs are shifted to higher energies, as a result of the increase of the average lateral confinement within the SiNW. However, with increasing $L_r$, the conduction band edge moves to lower energy, while the DoS is steadily degraded, hardly showing any fine structure for $L_r\geq 2.17$ nm. The lowest-lying electron states are indeed trapped deeper in energy in the largest sections of the nanowire [see Fig. (\ref{figure1})a]. The extendedness of the electron wave functions will ultimately determine the transport regime \cite{Drabold}.

\begin{figure}[!t]
	\centering
	\includegraphics[width=8.5cm]{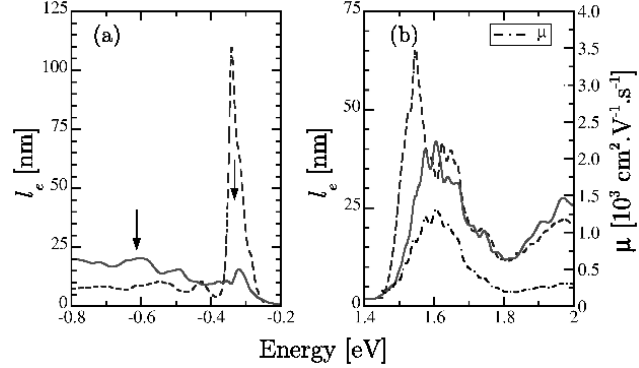}
	\caption{(color online). Mean free path of the holes (a) and electrons (b) as a function of energy, for $L_r=0.54$ nm (blue dashed line) and $L_r=2.17$ nm (red solid line). The electron mobility is also shown for $L_r=2.17$ nm (black dash-dotted line).}
	\label{figure2}
\end{figure}

\subsection{Elastic mean free paths and charge mobilities}

The mean free path of the electrons and holes are plotted as a function of energy in Fig.~\ref{figure2}, for the same two $L_r$'s as in Fig.~\ref{figure1}. The main features of the underlying band structure still show up at $L_r=0.54$ nm. Indeed, the electron mean free path reaches its maximum ($\ell_{e}\simeq70$ nm) between the first two VHs (single subband transport), shows a dip at the edge of the second subband, then further decreases above $E\simeq1.7$ eV due to enhanced interband scattering. The hole mean free path also exhibits a very sharp peak ($\ell_{e}\simeq 110$ nm) in the first subband, but becomes very short in the dense lower-lying subbands. However, at $L_r=2.17$ nm, the fine structure of $\ell_{e}$ can not be so easily related to the band structure of the ideal nanowire. The mean free path is almost reduced by half in the first electron ``subband'', while the peak on the valence band side is nearly five times smaller and is superseded by a broad feature at lower energies.

The charge carrier mobility $\mu$ is another key quantity for assessing device performances. It can be related to the Kubo conductivity $\sigma(E)=n(E)\mu(E)e$, where $n(E)$ is the charge density (per unit of length) and $e$ is the elementary charge \cite{triozon:121410,Ferry}. Alternatively $\mu(E)$ can be written as $e\tau_e(E)/m^*$, where $m^*$ is the effective mass of the carriers. These two expressions are fully equivalent under the assumptions of Drude's phenomenology with parabolic bands. However, the latter is impractical in SiNWs, because the bands have different effective masses (and might show significant non-parabolicity), and also because the SRD induces large changes in the electronic structure as evidenced previously. $\mu(E)$ is shown in Fig.~\ref{figure2}(b) for $L_r=2.17$ nm, and follows the same trends as $\ell_{e}$. 

\begin{figure}[!t]
	\centering
	\includegraphics[width=8.5cm]{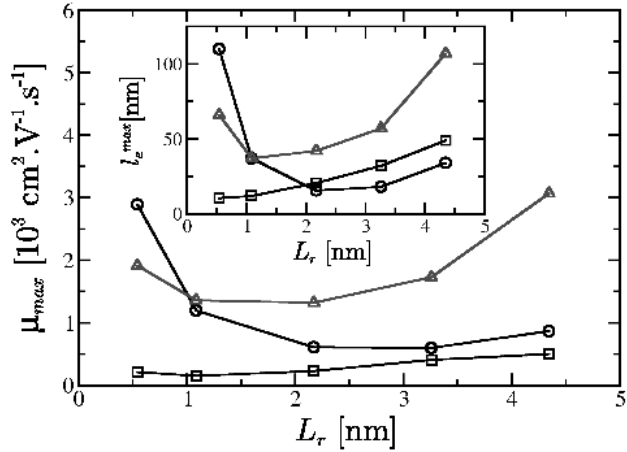}
	\caption{(color online). Maximum mobility for the electrons (red triangles) and holes (circles: first peak, squares: second peak) as a function of $L_r$. Inset: Maximum mean free path.}
	\label{figure3}
\end{figure}

The maximum mean free paths $\ell_{e}^{max}$ and mobilities $\mu_{max}$ are reported as a function of $L_r$ in Fig.~\ref{figure3} for both kinds of charge carriers. The heights of the first, sharp peak and of the second, broad peak at lower energies are both given for the holes [arrows in Fig. \ref{figure2}(b)]. The mobility is always maximum at the first peak, although the mean free path can be larger within the second one. The values of $\mu_{max}$, that are ranging from several hundreds to thousands of ${\rm cm}^{2}{\rm V}^{-1}{\rm s}^{-1}$, turn out to be in fair agreement with the experimental estimates for the most performant undoped semiconducting nanowires \cite{CuiY._nl025875l}. The mobility can nonetheless be much smaller around the band edges, as evidenced in Fig.~\ref{figure2}(b). Additionally, $\ell_{e}^{max}$ and $\mu_{max}$ show a minimum as a function of $L_r$. This suggests that electrons and holes are less sensitive to short length scale fluctuations of the SRD profile (at least in the first conduction and valence subbands) \cite{SRDlengthscale}. The carriers are much more efficiently scattered by the SRD at intermediate values of $L_r$. Finally, the surface of the nanowires becomes locally smooth again as $L_r$ is further increased, which enhances the mobility of the charge carriers far enough from the conduction and valence band edges.

\subsection{Conductance, quantum interference effects and localization regime}

For an in-depth analysis of transport properties in the coherent regime, the effects of quantum interferences effects on conductance scaling need to be analyzed in details. This can be conveniently achieved by using the Landauer-B\"uttiker transport framework \cite{Ferry}. This method is additionally well suited for describing transport between contacts connected to the nanowire, which becomes important when charge injection properties start to play an important. Properties such as transmission and local density of states are extracted from the Green's function and the self-energies of the contacts.

\begin{figure}[!t]
	\centering
	\includegraphics[width=8.5cm]{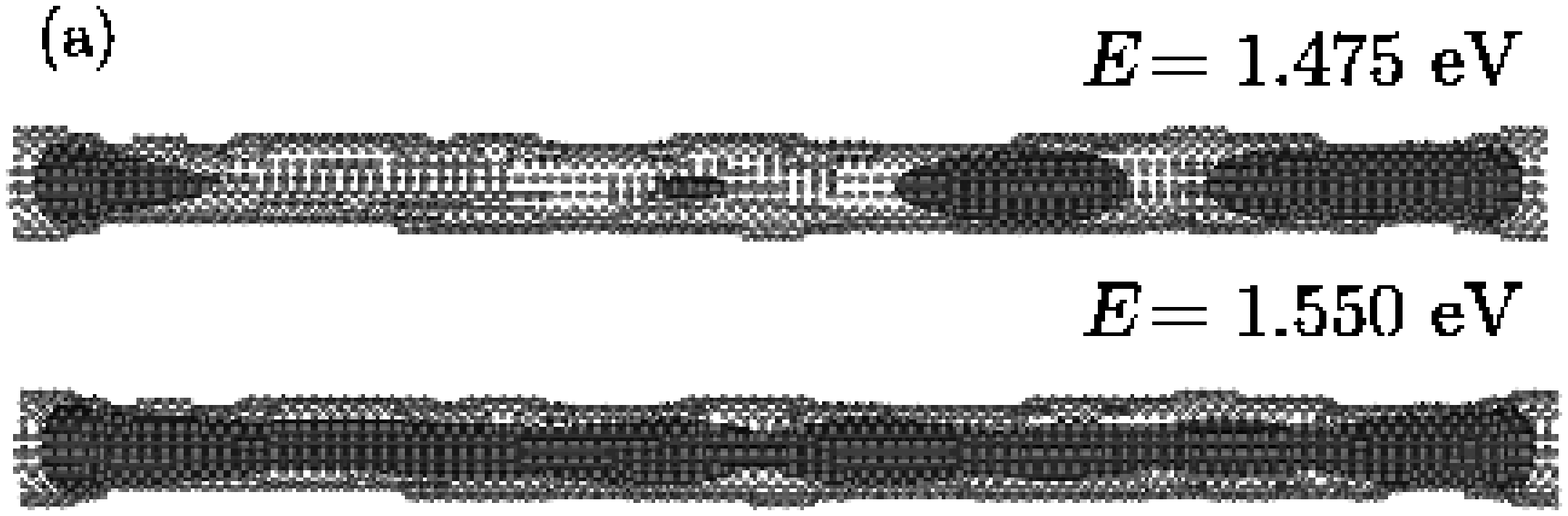}
	\includegraphics[width=8.5cm]{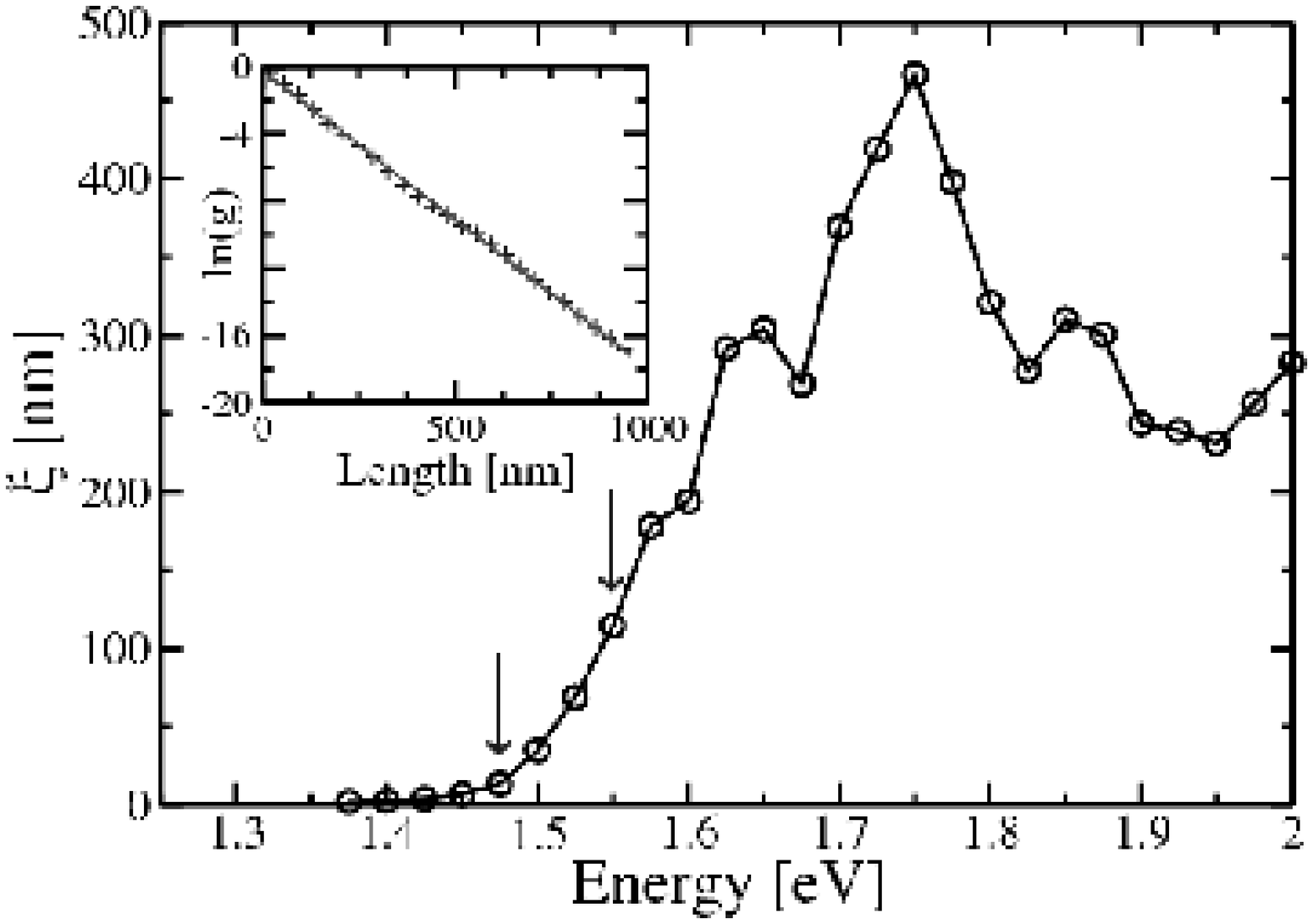}
	\caption{(color online). (a) Local density of states in a short nanowire with length $L\simeq 30$ nm at energies $E=1.475$ eV and $E=1.55$ eV [arrows in (b)]. (b) Localization length $\xi$ as a function of energy ($L_r = 2.17$ nm). (c) Linear fit (red solid line) to $\langle \ln g \rangle(L)$ (blue crosses) at $E=1.55$ eV.}
	\label{figure4}
\end{figure}

\begin{equation}
G= \frac{1}{(E-H)-\sum_{n} \Sigma_n},
\end{equation}
\begin{equation}
\Gamma_n(E) = i[\Sigma_n(E)-\Sigma_n^\dag(E)],
\end{equation}
\begin{equation}
T_{nm}(E) = Tr[\Gamma_n(E)G\Gamma_mG^\dag(E)],
\end{equation}
where $\Sigma_n(E)$ is the self-energy of contact $n$ and $T_{nm}(E)$ is the transmission between contact n and m. As the Green's function is calculated in segments using a Green's function decimation technique,\cite{Grosso89} it is an $O(R^6)$ method with regard to the nanowire radius, limiting practical calculations to $R \lesssim 1$ nm. The contacts are modelled as semi-infinite nanowire leads with no roughness and a radius of $R = R_0+0.2$ nm. A standard decimation technique is employed \cite{Troels,Grosso89} to compute the transmission through the SiNWs.

In absence of any SDR, the transport regime remains ballistic and the resulting computed Landauer conductance $G$ is quantized $G\simeq N_{\bot}G_0$ \cite{noteLB}, with $G_0=2e^2/h$ being the quantum of conductance, and $N_{\bot}$ gives the number of conducting channels. Conductance quantization has been experimentally observed in narrow constrictions (quantum point contacts) made on silicon \cite{QC1} and short silicon wires \cite{CuiY._nl025875l,QC2}. However, most of the available results performed on longer and wider lithographically defined Silicon nanowires evidence the strong contribution of disorder effects (due to wire width fluctuations or charged impurities) that produce quantum interferences, localization \cite{QC3} and charging effects \cite{QC4,QC5}. 

In this work, the effect of electron-electron interactions is neglected, and the focus is put on the transition from weak to strong localization. Indeed, in the absence of inelastic scattering, quantum interferences build up beyond the diffusive regime, lead to the localization of all wave functions in the zero-temperature limit \cite{Juanjo,Thouless,Beenakker}. The localization length $\xi(E)$ can be extracted from a scaling analysis of the conductance through $\langle \ln g(E) \rangle \sim -2L/\xi(E)$, where $g=G/G_0$ and $L$ is the length of the wire [Fig.~\ref{figure4}(c)]. $\xi(E)$ is plotted as a function of the conduction band energy in Fig.~\ref{figure4}(b) ($L_r = 2.17$ nm). The logarithm of $g$ was averaged (for each $L$) over 150 random SRD profiles is provided in the inset. $\xi$ ranges from a few nanometers close to the gap up to $\simeq 500$ nm at higher energies. The electrons indeed tend to localize in the largest sections of the nanowires near the conduction band edge [see Fig. (\ref{figure1})a]. As an other illustration, the local density of states is shown at two conduction band energies in Fig.~\ref{figure4}(a), for a short nanowire with length $L\simeq 30$ nm. The corresponding localization lengths are $\xi(1.475\hbox{ eV}) = 13.5$ nm and $\xi(1.55\hbox{ eV}) = 114.5$ nm.

\begin{figure}[!t]
	\centering
	\includegraphics[width=8.5cm]{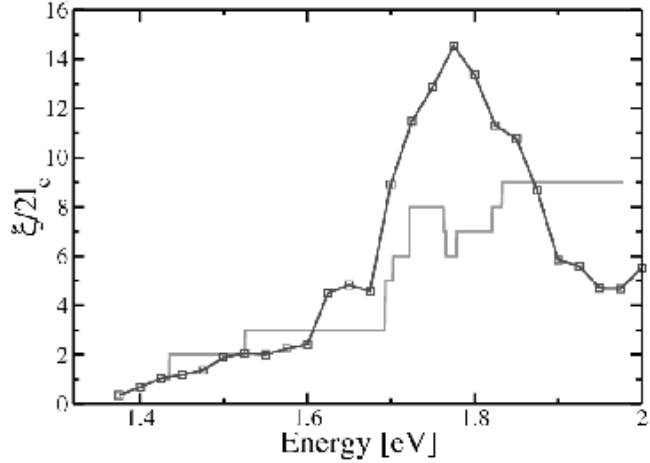}
	\caption{(color online). $\xi/(2\ell_{e})$ (red squares) and $N_{\bot}+1$ (green solid line) as a function of energy.}
	\label{figure5}
\end{figure}

In weakly disordered quasi-1D systems, the fundamental length scales $\xi$ and $\ell_e$ are shown to scale as $\xi\sim2(N_{\bot}+1)\ell_e$ \cite{Thouless,notele}. This relation was first established by Thouless \cite{Thouless} for strictly 1D systems, and further generalize with random matrix theory to weakly disordered quasi-1D systems with larger number of conducting channels \cite{Beenakker}. Recently, it was studied numerically for chemically doped disordered carbon nanotubes \cite{Avriller}. In our situation, the identification of a well-defined channel structure is however hindered in the SiNWs by the SRD-induced changes of the electronic structure. The comparison between $\xi/(2\ell_{e})$ and $N_{\bot}+1$ (deduced from the band structure of the ideal nanowire) is nonetheless instructive [Fig. \ref{figure5}]. As evidenced by our calculations, $\xi/(2\ell_{e})$ roughly scales as $N_{\bot}+1$ close to the conduction band edge, but the situation becomes much more complex at higher energies at which strong difference occur. Similar discrepancies were recently reported in simplified models of disordered quantum wires, but in the presence of a magnetic field \cite{Feist}.

%This possibly points to the limits of the weak disorder assumption underlying Thouless relation. 

\section{CONCLUSION}

In conclusion, some key transport length scales in disordered semiconducting nanowires have been investigated using an optimized real space order $N$ method combined with a recursive Green's function-based Landauer-B\"uttiker approach. The effectiveness of an atomistic-based surface roughness profile in limiting ballistic transport has been demonstrated, and the trends in the energy-dependent electron and hole mobilities, mean free paths and localization lengths have been discussed at a quantitative level. The limitations of the Thouless relationship in such complex disordered systems have also been pointed out. Studies focusing on the role of nanowires orientation, diameter, and other kind of disorder such as dopants \cite{niquet:165319,TROELS2}, surface defects, or traps in the oxide, deserve further consideration. Additionally, beyond the intrinsic effects of these various sources of scattering on transport length scales, their impact in SiNWs-based field effect devices should be investigated \cite{SiNWsELEC}.

\section{ACKNOWLEDGMENTS}

This work was supported by the French "Action Concert\'{e} Initiative" (ACI) "Transnanofils" and by the EU project No 015783 NODE. The calculations were performed at the CEA/CCRT supercomputing center. 

%\bibliography{Mobility}

\begin{thebibliography}{1}
\bibitem{noteH} All silicon atoms that end up with three hydrogen atoms as first nearest neighbors are also removed from the nanowire and replaced with a single hydrogen atom.
\bibitem{noteavg} The Kubo data were averaged over 5 to 10 realizations of the SRD
, the trace being computed with 8 random-phase states.
\bibitem{noteLB}
In the Landauer-B\"uttiker approach, the ideal and disordered SiNWs are connected to slightly larger SiNW electrodes, which gives rise to the Fabry-Perrot-like oscillations visible in the inset of Fig.~\ref{figure1}(c).
\bibitem{notele}
Our mean free path $\ell_{e}=v\tau_e$, consistent with kinetic theory, is half the mean free path defined, e.g., in Ref. \cite{Beenakker} -- hence the additional factor 2 in Thouless relation.
\end{thebibliography}

\end{document}